\documentclass[aps,prl,twocolumn,groupedaddress]{revtex4}

\usepackage{graphicx}
\usepackage{amsmath}
\usepackage{amscd}

\begin{document}

\title{Optical Spin Initialization and Non-Destructive Measurement in a Quantum Dot Molecule}

\author{Danny Kim}
\author{Sophia E. Economou}
\author{\c{S}tefan C. B\u{a}descu}
\author{Michael Scheibner}
\author{Allan S. Bracker}
\author{Mark Bashkansky}
\author{Thomas L. Reinecke}
\author{Daniel Gammon}
 \affiliation{Naval Research Laboratory, 4555 Overlook Ave, SW, Washington D.C., 20375}
\date{\today}

\begin{abstract}
The spin of an electron in a self-assembled InAs/GaAs quantum dot molecule is optically prepared and measured through the trion triplet states.  A longitudinal magnetic field is used to tune two of the trion states into resonance, forming a superposition state through asymmetric spin exchange. As a result, spin-flip Raman transitions can be used for optical spin initialization, while separate trion states enable cycling transitions for non-destructive measurement. With two-laser transmission spectroscopy we demonstrate both operations simultaneously, something not previously accomplished in a single quantum dot.
\end{abstract}



\maketitle

Initialization, coherent manipulation, and readout are the essential operations of quantum information processing.  The electron spin in a singly-charged InAs quantum dot can serve as a qubit for all-optical solid state quantum computing.  Localization of the electron greatly extends its spin coherence times \cite{kro04nat,gre07sci} and the spin can be addressed through an optically excited trion state \cite{ram08prl,ata06sci,xu07prl,ger08nat,xu08nph,ber08sci,wu07prl}.  In a transverse magnetic field, the trion and the two spin states of the electron form a 3-level "$\Lambda$" system that enables spin initialization \cite{ata06sci,xu07prl,ger08nat} and control \cite{xu08nph,ber08sci,wu07prl} through Raman transitions.  The transverse field turns on the normally forbidden transitions by breaking the axial symmetry of the system.  A major drawback is that this precludes the use of sensitive 2-level cycling transitions.  In a cycling transition measurement the system continues to return to the same spin eigenstate because of strict selection rules, and in this sense is non-destructive, as for example in the case of ion qubits \cite{mon95prl}.  Non-destructive readout is necessary for error correction during quantum calculations.  Also, higher measurement sensitivity is possible enabling single-shot readout.  Excited orbitals in single dots, which contain both singlet and triplet states, could be used except that they suffer from fast nonradiative relaxation \cite{war05prl}.  Thus spin initialization and manipulation are incompatible with non-destructive cycling readout in single dots.\\
\begin{figure}
\begin{center}
\includegraphics[width=8cm,keepaspectratio=true]{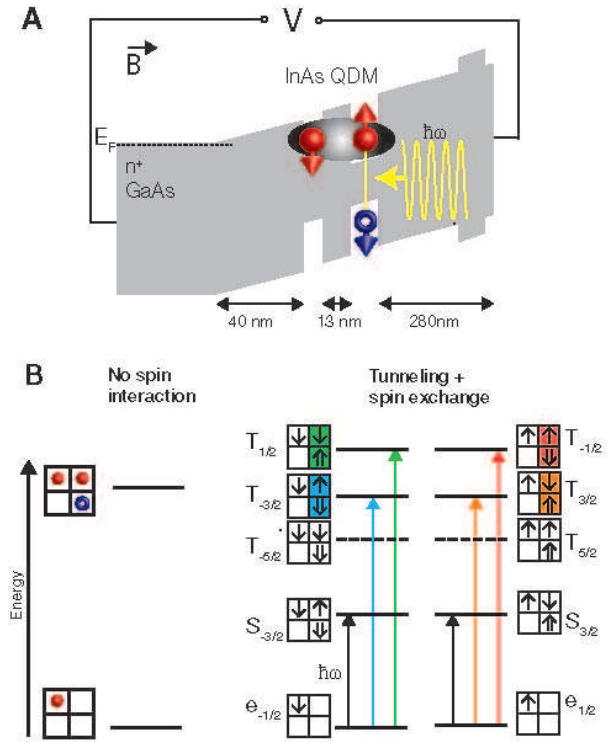}
\end{center}\caption{(Color online). Transmission spectroscopy of a QD molecular trion.  (a) Schematic diagram of the device structure showing the two electrons (solid circle) and hole (open circle) of the trion.  (b) Energy level diagram of both the electron and trion states.  Arrows indicate allowed optical transitions. The levels are labelled by the spin configurations of the states and their total spin projections.}\label{device}
\end{figure}
In this Letter, we overcome this fundamental limitation by using a pair of quantum dots that are coupled through coherent tunnelling \cite{sti06sci,kre06prl,sch07prb}. Optical excitation of one dot is used to initialize and to readout the spin state of an electron in the other dot through exchange interactions.  The unique energy level structure in coupled dots eliminates the need for a transverse field.  Instead a longitudinal field is used to tune two trion states into resonance such that a small asymmetric exchange interaction permits a spin-flip Raman process.  At the same time, other states maintain good selection rules and are used for cycling transition measurement.  Overall, the singly-charged coupled quantum dot forms a ``\textsf{W}'' energy level system, which is comprised of a $\Lambda$ system and two two-level cycling transitions.    With this versatile new qubit, we are now able to demonstrate simultaneous spin initialization and non-destructive readout.

Our qubit is realized in two vertically stacked InAs self-assembled quantum dots separated by a 13~nm GaAs barrier and electrically biased in a diode structure so that a single electron resides in the bottom QD [Fig.\ref{device}(a)].  Single molecules for optical study are isolated by 1~$\mu$m diameter apertures in an Al shadow mask. The sample bias is modulated with a square-wave voltage of 50~mV peak-to-peak at 10~KHz. Lock-in techniques are used to measure the changes in the transmitted laser, which is linearly polarized and focused to a $\sim$2~$\mu$m spot.

To initialize and readout the spin of this electron an additional electron-hole pair is optically excited in the top QD.  The structure is designed so that the electrons can tunnel, whereas the hole cannot \cite{bra06apl,kre05prl}.  The tunneling of the two electrons results in spin states in which a singlet is separated in energy from three triplet states [Fig. \ref{device}(b)] \cite{sch07prb,dot06prl}.  The triplet states are further split through the $e$-$h$ exchange interaction with the hole spin. We first present the energy level structure of this system and then show how it results in a \textsf{W} diagram that can perform simultaneous optical spin initialization and measurement.
\begin{figure}
\begin{center}
\includegraphics[width=8.6cm,keepaspectratio=true]{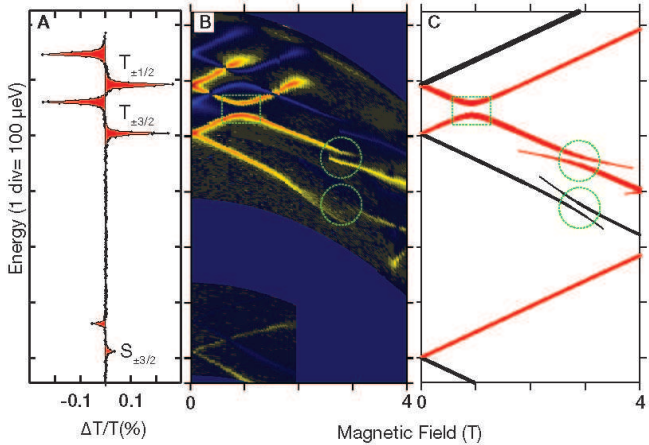}
\end{center}\caption{(Color online). (a) Transmission spectra of the molecular trion at fixed bias.  The negative peaks in the spectra arise from the voltage modulation technique and are just replicas of the positive peaks\cite{hog04prl}. (B) Intensity plot of the transmission spectra as a function of longitudinal magnetic field. A diamagnetic contribution to the energy (10.8~$\mu$eV/T$^2$) has been subtracted.  Energy anticrossings are observed at B=1~T (square) and 2.8~T (circles).  (C) Calculated transition spectra in which line thicknesses are proportional to the oscillator strength. The black and red lines correspond to transitions from electron spin down and up, respectively.}\label{fan}
\end{figure}

\begin{figure}
\begin{center}
\includegraphics[width=8.6cm,keepaspectratio=true]{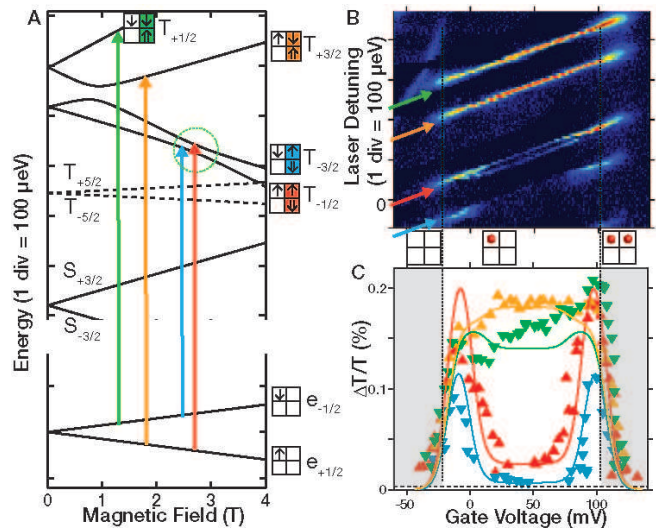}
\end{center}\caption{(Color online). (a) Calculated state energies as a function of magnetic field at fixed voltage (50~mV).  (b) One-laser transmission spectra as a function of voltage.  Intensity plot of the four transitions in the \textsf{W} level diagram across the one-electron stability plateau. B=2.75~T and laser power is 3~$\mu$W and linearly polarized.  (c) Peak intensities of the four measured (symbols) and calculated (lines) transitions.  The ground state charge configuration is shown between the two plots, and the grey shaded areas denote regions of bias space where the molecule is charged with 0 and 2 electrons.}\label{plateau}
\end{figure}

Two of the triplets are shown in the optical transmission spectra in Fig. \ref{fan}(a).  The third triplet is optically forbidden due to selection rules.  The transitions split into Zeeman components [Fig. \ref{fan}(b)] with an applied longitudinal magnetic field.  The optical spectra arises from transitions from the spin states of the resident electron to the spin states of the trion.  The red lines in Fig. \ref{fan}(c) indicate a transition that originates from the spin $+\frac{1}{2}$ state of the resident electron and the black from the spin $-\frac{1}{2}$ state.  We use the Hamiltonian in Ref. \cite{sch07prb} and augment it with additional terms for asymmetric exchange \cite{kav04prb,bad05prb} as described below.  A longitudinal field (Faraday geometry) maintains the zero-field selection rules [see Fig \ref{device}(b)], unlike the case of a magnetic field applied in the transverse direction (Voigt geometry). The calculated transition spectra shown in Fig. \ref{fan}(c) are in excellent agreement with the measured spectra of Fig. \ref{fan}(b).  Fig. \ref{fan}(c) is the difference between the calculated state energies (plotted in Fig. \ref{plateau}(a)) using the selection rules depicted in Fig. \ref{device}(b).

In the magnetic field data of Fig. \ref{fan}(b) it is seen that where transition energies would cross there are avoided crossings, or anticrossings.  The first anticrossing at B=1~T [square in Fig. \ref{fan}] corresponds to a coupling between two basis states $T_{-\frac{1}{2}}=\left(^{\uparrow\,\,\,\,\,\uparrow}_{0\,\,\,\,\,\Downarrow}\right)_T$  and $T_{+\frac{3}{2}}=\left(^{\uparrow\,\,\,\,\,\downarrow}_{0\,\,\,\,\,\Uparrow}\right)_T$  , which differ both in a hole and in an electron spin projection in the top dot.  This coupling can arise from asymmetric (sometimes called anisotropic) $e$-$h$ exchange.  This anticrossing is analogous to the fine-structure splitting normally seen in the neutral exciton spectra in a single dot and arises from the same origin \cite{gam96prl}. As expected, the polarization selection rules change from circular to linear at this anticrossing point.

The key to our spin initialization method is the second anticrossing that occurs at B=2.8~T between the trion states $T_{-\frac{1}{2}}=\left(^{\uparrow\,\,\,\,\,\uparrow}_{0\,\,\,\,\,\Downarrow}\right)_T$ (red) and $T_{-\frac{3}{2}}=\left(^{\downarrow\,\,\,\,\,\uparrow}_{0\,\,\,\,\,\Downarrow}\right)_T$ (blue) with a magnitude of $\delta_{ee}\approx15$~$\mu$eV (circle in Fig. \ref{plateau}(a)).  At the magnetic field where the two trion states anticross a small asymmetric exchange contribution becomes dominant and becomes directly measureable through the magnitude of the anticrossing energies. These two trion states differ by the orientation of one electron spin, and at the anticrossing the state becomes a coherent superposition of both trion states.  The superposition state has strong optical transition strength with both spin states of the resident electron, which turns on the normally optically inactive transitions in the region of the anticrossing.   This explains the two observable anticrossings in the transition spectra (Fig \ref{fan}(b) and \ref{fan}(c)).   The superposition state and the two electron spin ground states form a $\Lambda$ system.  Thus, spin-flip Raman transitions can be performed through these superposition states.

We first present the single-laser transmission spectra as a function of gate voltage in Fig. \ref{plateau}(b).  In Fig. \ref{plateau}(c), the intensities of the spectral lines from Fig. \ref{plateau}(b) are plotted.  The two lower lines in Fig. \ref{plateau}(c) correspond to the initialization transitions, and they show a sharp drop in intensity in the middle of the bias range: this is a signature of optical spin pumping \cite{ata06sci,xu07prl,ger08nat}.  When the electron spin is optically excited to the trion superposition states, it can recombine back to the other electron spin state, where it is shelved until it relaxes.  Because the initial spin eigenstate is no longer populated, the $\tfrac{\Delta T}{T}$ signal is reduced.  This pumping rate is fast ($\sim$ 1~ns) because both branches of the $\Lambda$ transitions have large transition strength as a result of the strong exchange-induced coupling between the trion states.  At the edges of the bias range, optical pumping is suppressed by rapid co-tunneling of electrons between the quantum dot molecule and the doped GaAs layer \cite{smi05prl,dre08prb}.

In contrast, transitions to the unmixed triplet states $T_{+\frac{1}{2}}=\left(^{\downarrow\,\,\,\,\,\downarrow}_{0\,\,\,\,\,\Uparrow}\right)_T$  and $T_{+\frac{3}{2}}=\left(^{\uparrow\,\,\,\,\,\downarrow}_{0\,\,\,\,\,\Uparrow}\right)_T$  (green and orange, respectively) remain intense over the single electron stability plateau [Fig. \ref{plateau}(b)] and show no signs of optical pumping.  This is due to the fact that each of these trion states ($T_{+\frac{1}{2}}, T_{+\frac{3}{2}}$) couple optically to only one electron spin state, so that the original electron spin eigenstate is recovered by spontaneous emission.  In particular, the transition involving the $T_{+\frac{1}{2}}$ trion state:
\begin{equation}
\begin{CD}
\left(^{\downarrow\,\,\,\,\,\,0}_{0\,\,\,\,\,\,0}
\right)@>\hbar\omega>>\left(^{\downarrow\,\,\,\,\,\downarrow}_{0\,\,\,\,\,\Uparrow}
\right)_T@>\hbar\omega>>\left(^{\downarrow\,\,\,\,\,\,0}_{0\,\,\,\,\,\,0}\right)
\end{CD}
\end{equation}
is robust against heavy/light hole mixing, which is known to break the selection rules in single dots.  These cycling transitions can be performed repeatedly to provide efficient, non-destructive measurement of the spin eigenstate \cite{mon95prl}.

The calculated lines in Fig. \ref{plateau}(c) are the steady-state solutions to the optical Bloch equations combined with an expression for the co-tunneling rate from earlier studies of single dots \cite{smi05prl,dre08prb,kro08prb}.  Good agreement was found using a transition dipole of 25~D and spontaneous emission rate of (500~ps)$^{-1}$ for the $T_{+\frac{1}{2}}$ transition and similar values for the other transitions.
\begin{figure}
\begin{center}
\includegraphics[width=8.6cm,keepaspectratio=true]{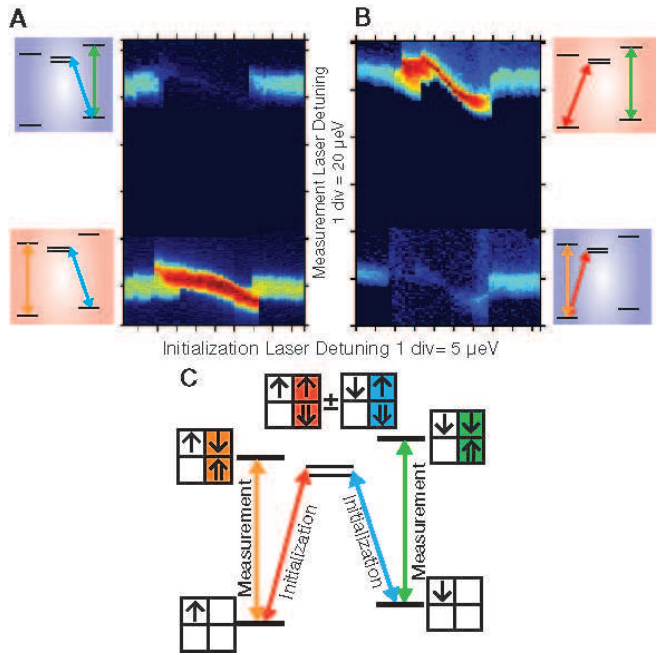}
\end{center}\caption{(Color online) Simultaneous spin initialization and measurement with two lasers at B=2.75~T.  The initialization and measurement lasers are at 3.5~$~\mu$W and 3~$~\mu$W, respectively. (a)-(b) Intensity plot of measurement laser transmission (orange and green arrows) as a function of the initialization laser frequency (red and blue arrows) for initialization laser resonant first with spin down (a) and then with spin up (b). The two traces in each plot show the intensities of the two measurement transitions for the same initialization laser. The two lasers are cross linearly polarized. A polarization analyzer before the detector transmits only the measurement laser. (c) ``\textsf{W}'' energy level diagram showing the spin configuration of each level.}\label{twolaser}
\end{figure}

In total the four transitions connected to the two ground spin states of the electron define a \textsf{W} level diagram [Fig. \ref{twolaser}(c)] that enables simultaneous spin initialization and measurement.  To demonstrate this we performed a two-laser transmission experiment [Fig. \ref{twolaser}].  The initialization laser is scanned through resonance with the superposition doublet while the measurement laser is scanned through both measurement transitions [see Fig. \ref{twolaser}(a) and \ref{twolaser}(b)].  When the initialization laser is resonant with the transition from the $+\frac{1}{2}$ spin state to either component of the superposition doublet (red arrow), the electron spin is pumped from $+\frac{1}{2}$  to $-\frac{1}{2}$.  As a result, the intensity of the green measurement transition is enhanced, while the orange measurement is suppressed [as shown in Fig. \ref{twolaser}(b)].  The reverse is obtained when pumping from the $-\frac{1}{2}$ spin state (blue arrow) as shown in Fig. \ref{twolaser}(a). The difference between the intensity of the enhancement and suppression of the measurement transitions reflects the population difference.  Using the optical Bloch equations we obtain a spin polarization $\frac{n_\uparrow-n_\downarrow}{n_\uparrow+n_\downarrow}$ $\approx$96\% at saturation.  This value is somewhat lower than that obtained previously in single dots, and probably results from somewhat larger optical linewidths ($\sim$2~GHz here as compared to 0.4~GHz in Ref. \cite{ata06sci}).  The optical linewidth originates from spectral wandering that likely arises from charge fluctuations in the surrounding material, and from our simulations can account for the reduction in pumping fidelity.

We now return to the origin of the anticrossings.   The anticrossing observed at B=2.8~T corresponds to a coupling between two basis states that differ only in a single electron spin projection.  We find that this coupling can arise from spin-orbit interaction.  This leads to an electron exchange between dots that is accompanied by a spin flip.  The spin-flip Raman process that drives the spin pumping is described by:
\begin{equation}
\begin{CD}
\left(^{\uparrow\,\,\,\,\,\,0}_{0\,\,\,\,\,\,0}
\right)@>\hbar\omega>>\left(^{\uparrow\,\,\,\,\,\uparrow}_{0\,\,\,\,\,\Downarrow}
\right)_T@>\beta^{ee}_->>\left(^{\uparrow\downarrow\,\,0}_{0\,\,\,\,\,\Downarrow}
\right)_S@>t_e>>\ldots\\
\ldots\left(^{\downarrow\,\,\,\,\,\uparrow}_{0\,\,\,\,\,\Downarrow}
\right)_S@>\beta^{ee}_z>>\left(^{\downarrow\,\,\,\,\,\uparrow}_{0\,\,\,\,\,\Downarrow}
\right)_T@>\hbar\omega>>\left(^{\downarrow\,\,\,\,\,\,0}_{0\,\,\,\,\,\,0}
\right)
\end{CD}
\end{equation}
The basis states are defined as in Ref. \cite{sch07prb}. The interaction terms are the asymmetric exchange $\beta^{ee}_{-}=(h^{e_1}_{so+}-h^{e_2}_{so+})(\sigma^{e_1}_--\sigma^{e_2}_-)$, which arises from the spin-orbit interaction $h^e_{so+}$; spin conserving tunnelling $t_e$; and the axially-symmetric (sometimes called isotropic) exchange $\beta^{ee}_z=(\Delta^{e_1h}_0-\Delta^{e_2h}_0)(\sigma^{e_1}_z-\sigma^{e_2}_z)\sigma^h_z$ between the electrons and a hole localized on one dot \cite{sch07prb}.

Using a perturbation analysis, an effective asymmetric exchange interaction between the two triplet states is found to be:
\begin{equation}
\delta^{ee}\approx\frac{\langle L_e|h^e_{so}|U_e\rangle}{t_e}\langle U_eU_h|\Delta^{eh}_0|U_eU_h\rangle
\end{equation}
$L$ and $U$ are lower and upper dot orbitals, respectively.  The tunneling rate ($t_e$=$850$~$\mu$eV) and the $e$-$h$ exchange energy ($\Delta^{eh}_0$=$130$~$\mu$eV), are known from measurement.  The spin orbit term $h^e_{so}\approx$95~$\mu$eV can be determined from the magnitude of the anticrossing energy ($\delta_{ee}$=$15$~$\mu$eV). We compare this value with a microscopic calculation in which we introduce a structural asymmetry by laterally displacing the two dots.  The spin-orbit interaction is the sum of Dresselhaus and Rashba couplings, $h^e_{so+}$=$(\alpha_D+i\alpha_R)p_+$.  We find that a lateral offset of $1-2$~nm is sufficient to account for the magnitude of the anti-crossing energy. Such an offset is physically reasonable \cite{sch08nph}, and we conclude that the spin-orbit-interaction is a viable origin for the mixing of the two triplet states at the anti-crossing point.

The spin-orbit interaction combined with lateral asymmetry can account for the magnitude of the anti-crossing energy, but by itself cannot account for several lineshape anomalies observed in Fig. \ref{twolaser}. Close inspection shows an energy shift as large as $15~\mu$eV and changes in linewidth.  This type of behavior is indicative of hyperfine interactions of the electron spin with the nuclear spin \cite{bro96prb,bra08sst,bra06prb,mal07prb,kor07prl,tar07prl}. At the anti-crossing point where the energy required for an electron-spin flip becomes small it is possible for the electron-nuclear spin flip process to become more efficient, leading to significant nuclear spin polarization. Moreover, the laser can induce a positive feedback process in which a spontaneous nuclear spin polarization is amplified and stabilized by the optical transition \cite{kor07prl}. A full treatment is beyond the scope of the present work and would likely involve a model that incorporates both spin-orbit-induced and hyperfine-induced electron spin flip processes.

We have demonstrated simultaneous initialization and non-destructive readout using resonant transmission spectroscopy. The readout method in the \textsf{W} energy diagram is not specific to a particular technique and resonance fluorescence \cite{mul07prl,vam08nat} or Faraday rotation \cite{ber06sci,ata07nph}, could also be used. Finally we note that the $\Lambda$ transitions used for spin initialization can also be used for coherent spin control, in order to set up a coherent superposition of the electron spin states  (for example, coherent population trapping in frequency domain \cite{xu08nph}, and coherent spin rotations in time domain \cite{wu07prl,ber08sci}).

\begin{acknowledgments}
We thank M.F. Doty and V.L. Korenev for illuminating discussions.  Partial funding was provided by NSA/ARO and ONR.  S.E.E. was supported by NRC/NRL.
\end{acknowledgments}

\end{document}